\documentstyle[aps,multicol,epsfig,eqsecnum]{revtex}
\begin{document}
\draft

\title{Scaling properties of the Penna model}

\author{E. Brigatti $^{\dag,}\thanks{E-mail address: edgardo@cbpf.br}$,
         J.S. S\'a Martins $^{\star}$,   
         and I. Roditi $^{\dag}$ 
        }  
\address{$\dag$Centro Brasileiro de Pesquisas F\'{\i}sicas, Rua Dr. Xavier 
  Sigaud 150, 22290-180, Rio de Janeiro, RJ, Brasil} 
\address{$\star$Instituto de F\'{\i}sica, Universidade Federal Fluminense, 
  Campus da Praia Vermelha, 24210-340, Niter\'oi, RJ, Brasil}
  
 \maketitle
 \widetext

\begin{abstract}
We investigate the scaling properties of the Penna model, which has become 
a popular tool for the study of population dynamics and evolutionary problems in 
recent years. We find that the model generates a normalised age distribution for which a 
simple scaling rule is proposed, that is able to reproduce qualitative features for 
all genome sizes.
\end{abstract}


\begin{multicols}{2}    
    
\section{Introduction}

In the last years, the usage of computational models has turned into a major trend 
in the discussion of problems in population dynamics and evolutionary theory. One of 
the reasons for this choice is undoubtedly the lack of substantial amounts of 
observational data on the dynamics of such systems; another, is the ability of 
computational models in mapping the dynamics of a non-Hamiltonian system into a set 
of simple rules of interaction between the large number of its individual 
constituents. Simulations of populations evolving under this set of rules serve as 
grounding test for the theoretical ideas that inspired them. The outcome of these 
simulations can then provide support for the role played by each particular 
conjecture, thus helping the theorist in providing guidelines for her or his work.

Statistical physicists have pioneered this effort, and their toolbox has proven its 
value in a number of different problems - see Ref. \cite{reviews} for recent 
reviews. Among the different models that have been used by physicists in the field, 
one stands out for its popularity. The Penna model \cite{penna} owes its leading 
role to a number of successes, and has further more managed to attract the attention 
of some theoretical biologists \cite{cebrat}. Despite - or perhaps because of - its 
simplicity, it has shown enough power to unravel the key  factors involved in such 
phenomena as the catastrophic senescence of semelparous species, female menopause 
and species branching under ecological pressure.

In the Penna model, individuals are represented by their genome, mapped onto one 
(haploid version) or two (diploid version) bit-strings. The standard genome used in 
the Penna model is 32-bit long, by no other reason than to turn it easy to 
implement on 32-bit word processors. In a study of the mortality data of the German 
population with the Penna model, genomes were represented by 128-bit long strings 
\cite{PS_01}. There, it was shown that it is possible to compare results for two 
different genome sizes by effecting a rescaling of some parameters. This result 
motivated the search for scaling in general, but a first proposal in this 
direction \cite{KM_01} was not conclusive. That computer simulation used an asexual 
model with a classical Verhulst factor and tried to compare directly results with 
different rescaled parameters. Another version of the asexual Penna model, 
continuous in time and using a real-valued genotype, was also object of a similar 
analysis \cite{RA_01}, but its results are not easily mapped onto the usual 
discrete version. Our approach, as can be seen in the following, is quite different. 

\section{A proposal for scaling analysis}

We are interested in studying the sensitivity of the Penna model for diploid 
individuals, that use sex for reproduction, with respect to the number of bits used 
in the implementation of the age-structured genetic load, and we focus on the 
analysis of the age distribution of the population.

In the Penna model, each position (locus) of the genome may contain a bit set to 
$1$ (harmful allele) or $0$. The passage of time in an individual's life triggers 
the activation of one further allele in the sequentially read bit-string. The 
amount of active harmful alleles determine the genetic death of the individual when 
it reaches some pre-determined threshold value. An individual may also die because 
of intra-specific competition for resources of the environment, and this is usually 
represented by a density-dependent mean-field death probability, called the 
Verhulst factor. A Fortran code that simulates the model, and was the basis for our 
own simulations, can be found in Ref. \cite{book}. Because the genome is 
age-structured, from a physical point of view we are studying the properties of 
the model's temporal scaling. The biological aspect of our analysis is to provide 
an answer to the question whether the model shows dependence on the genome size.

As a first step, we have looked for the model version more suited to the analysis 
and chose a variation of the version with a Verhulst factor that operates only on 
the first time step of an individual's life \cite{JSM_01}. Since this usage of the 
Verhulst factor is equivalent to setting the reproduction probability dependent on 
the population size, we made it explicitly by letting a female give birth with a 
probability given by $1$ minus the Verhulst factor. With this choice, we are able to 
control the population in a way that is not dependent on the genome length, due to 
the fact that a living individual never feels the effect of an external death 
factor. In the version with the usual Verhulst strategy, each living individual 
has an yearly probability to die, and the effective non-genetic death probability is 
trivially dependent on the genome size.

\begin{figure}
\begin{center}
\vspace*{0.8cm}
\resizebox{0.4\textwidth}{!}{\includegraphics{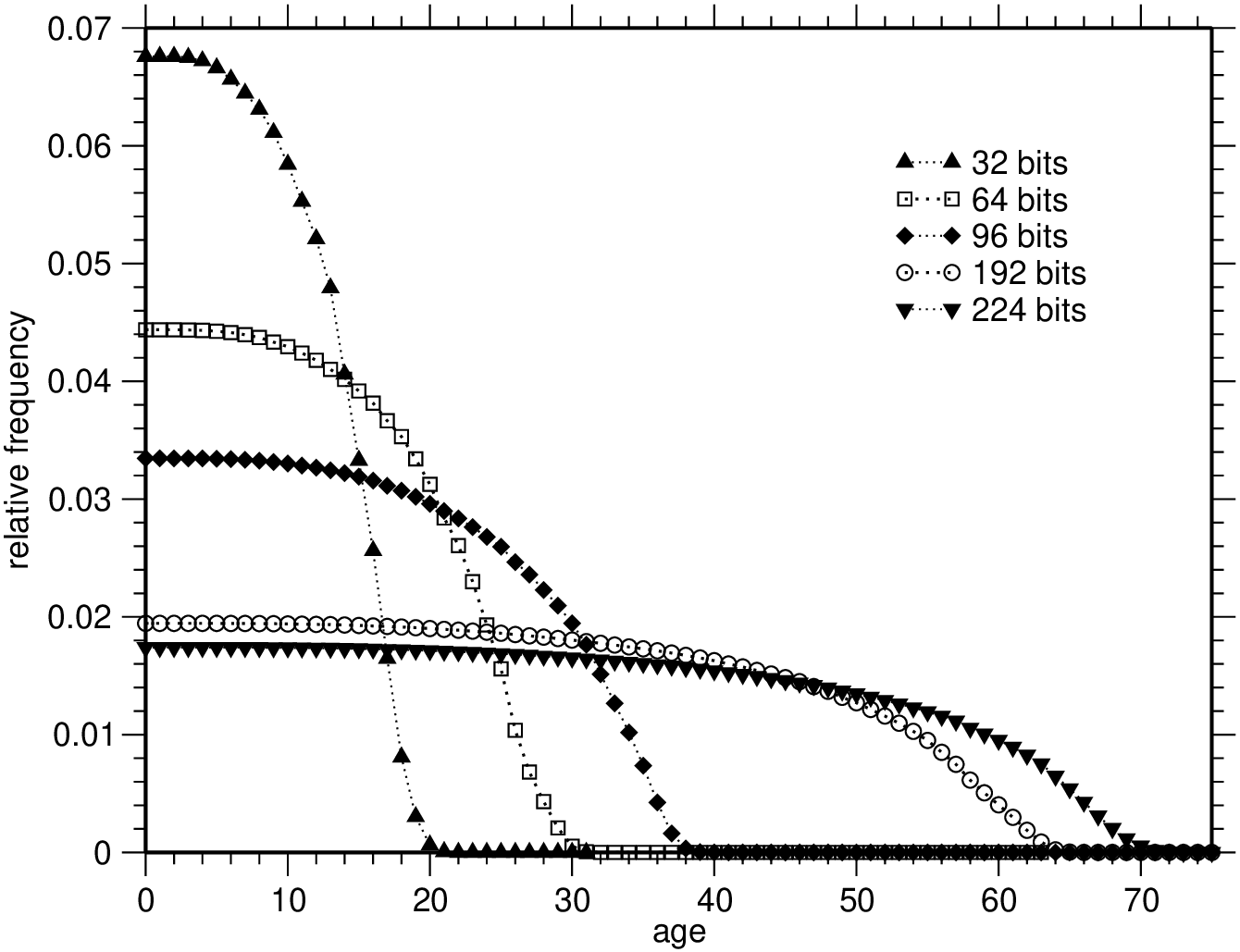}} 
\vspace*{0.4cm}
\end{center}
\caption{\small Age distribution of the population for a 32, 64, 96, 192 and 224 
bits models. The parameter used in the simulations are: the Verhulst parameter 
(400000), the initial population (1000), the minimum reproduction age (8), the 
number of offspring per mating season (4), the threshold value for harmful diseases 
(3), the number of mutations added at birth per bit string (1) and the number of 
dominant loci (6). We have averaged over the last 1000 steps of 10 different 
realisations in each case, after all statistical distributions could be confidently 
considered as stationary (simulations end after between 50000 and 200000 steps, 
depending on the size of the bit strings).}
\label{fig_Age} 
\end{figure}

In Fig. \ref{fig_Age} we show the age distribution of the population for various 
simulations of the model, differing by the number of bits in the genome. 
In the figure we are showing just the normalised values of the age distributions, 
forgetting the fact that the population grows with genomes' elongation. 
The first model uses a string of 32 bits. Because 32 is a natural unit for these computational 
studies running on 32-bit word processors, the other bit strings are chosen with 
sizes multiple of this number: 64, 96, 192 and 224 bits. In all the simulations 
performed, the parameters that control the number of dominant loci, the age at which 
reproduction starts, the number of offspring and mutations in each generation, and 
the threshold of harmful mutations are exactly the same. It is also relevant to notice that the 
cross-over frequency during gamete production is always one in each of these simulations.
The end of reproduction age is different for each genome length, in each case set to correspond to the 
maximum allowable age of the individuals (32, 64, 96, 192 or 224). In fact, our simulation 
conditions permit autosustaining populations that at equilibrium are not very sensitive to 
a change of this parameter. So, our choice does not introduce any undesirable asymmetry.
We can see that the distributions, although qualitatively similar, undergo a visible 
differentiation. We were led by their aspect to look for a scaling law that gives a 
relation between two different ages ($t_1$ and $t_2$) at which the integral of two 
distributions corresponding to different genome sizes ($\rho_1(t)$ and $\rho_2(t)$) 
reach the same population value. Formally, we search for a temporal rescaling 
$t_2 = F(t_1)$ that solves the equation 

\begin{equation}
\int_{0}^{t_1} dx \,\rho_1(x) = \int_{0}^{t_2} dx\,\rho_2(x)
\end{equation}

The solution turns out to be a very simple linear relation. 
In all cases, the integral of the distribution, as a function of its upper limit, 
starts with a linear growth, at ages where the distribution is essentially constant, 
and ends with a saturation, at the end of the lifespan of the population.
This behaviour suggests that a linear relation between the time values may satisfy 
the integral equality: If $y_i=a(i)+b(i)t_i$ is a regression of the linear part of 
the integral function of the distribution $\rho_i(t)$, $y_i=y_j$ leads to the 
relation we are looking for:

\begin{equation}
t_j = (b(i)/b(j)) \cdot t_i+ (\frac{a(i)-a(j)}{b(j)}) = c(i,j) \cdot t_i + d(i,j)
\label{lin2}
\end{equation}

\begin{figure}
\begin{center}
\vspace*{0.8cm}
\resizebox{0.4\textwidth}{!}{\includegraphics{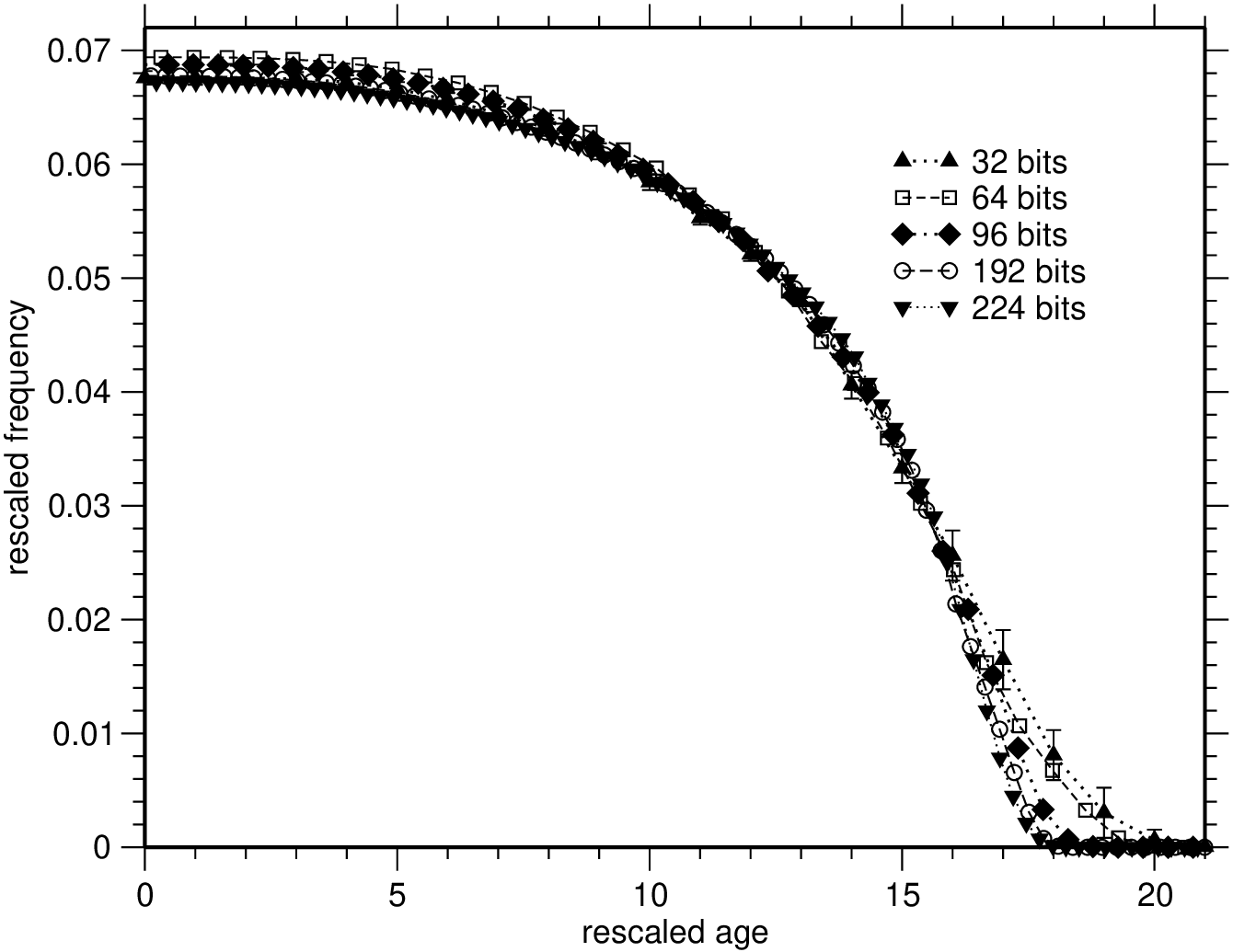} } 
\vspace*{0.4cm} 
\end{center}
\caption{\small By transforming the time scale of the distributions for genome 
sizes that are multiples of 32 using the inverse of the transformations with 
coefficients given in Table \ref{indices}, and then normalising them, it is 
possible to approach the 32-bits simulation from any of the others.}
\label{fig_RSca}
\end{figure}

Each index, $i$ or $j$, is defined as the bit string size divided by $32$. We can 
determine the coefficients of this rescaling relation by performing the regression 
of each integral function and using the above derived formulae to compute 
$c(i,j)$ and $d(i,j)$. For simplicity of notation, we omit the first index if it is 
equal to $1$. Table \ref{indices} shows results for these coefficients for some 
values of $j$ and for $i=1$.
These simple transformation relations allow both a proper rescaling of the full 
integral functions, and not only of its linear part, and also a rescaling of the 
age distributions. In fact, if we perform the inverse of the time transformations 
with the coefficients in Table \ref{indices} and then renormalise the rescaled 
functions, we obtain results that are close to the 32-bits distribution from all 
the others (See Fig. \ref{fig_RSca}).\\

From the coefficients listed in Table \ref{indices} it is possible to suggest a 
simple approximation for the slopes of the time rescaling transformations: 

\begin{equation}
c(j) \simeq [1 + 0.5 \cdot (j-1)].
\label{law}
\end{equation}

These coefficients are physically related to the model's temporal scaling, as 
already pointed out. A similar relation also holds for the terms $d(j)$, which are 
obtained as a difference between the constant terms of the regressions of the 
integral functions rescaled by a slope, and thus depend on the values of the age 
distributions at zero age.

We now compare the mortality functions, derived from the age distributions by the 
equation
\begin{equation}
f(a) = \log(\rho(a)/\rho(a+1)),
\end{equation}
where $\rho(a)$ is the value of the distribution at age $a$. 

\begin{figure} 
\begin{center}
\vspace*{0.8cm}
\resizebox{0.4\textwidth}{!}{\includegraphics{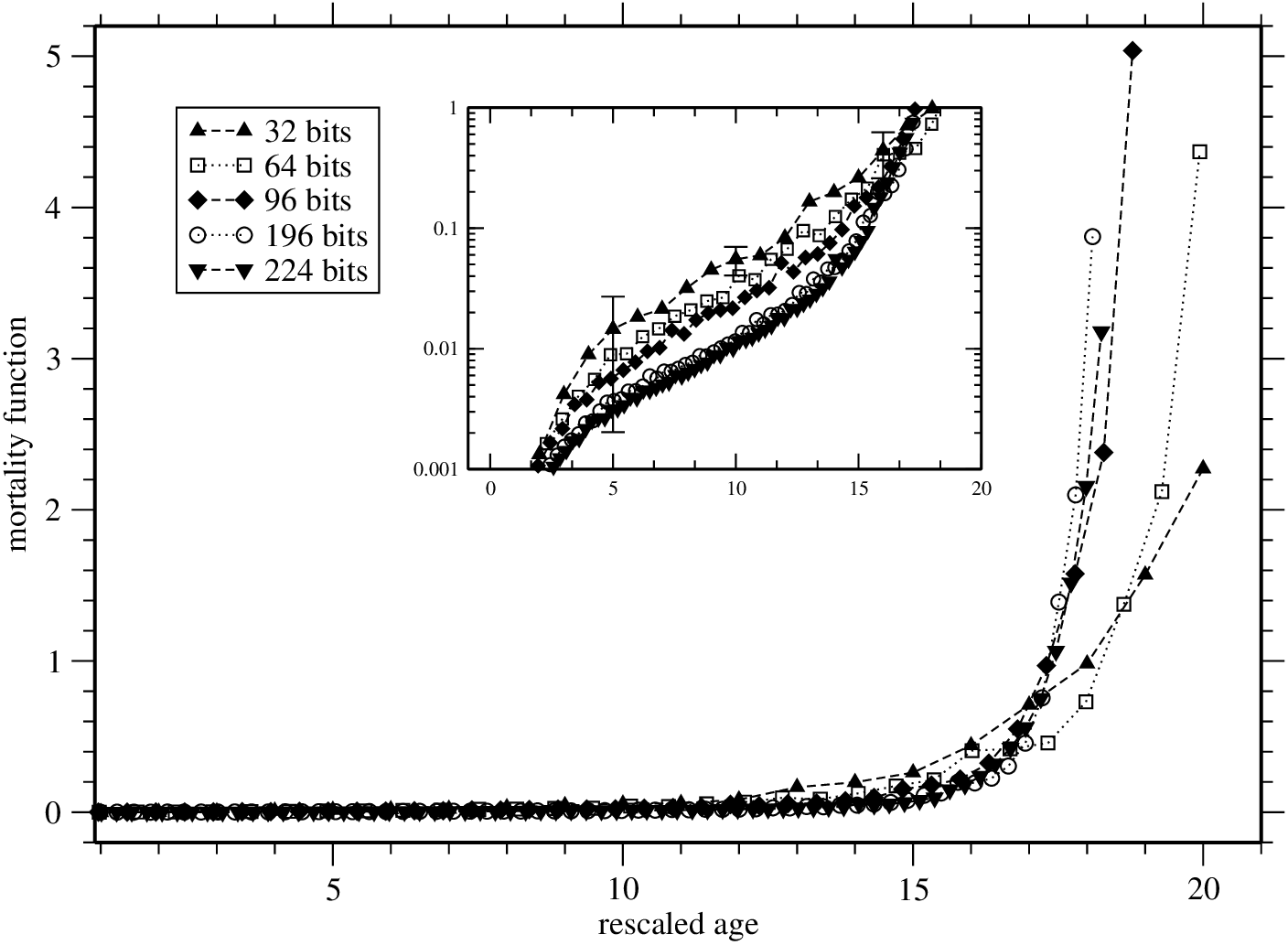} }
\vspace*{0.4cm}
\end{center}
\caption{\small The mortality functions computed from the rescaled age 
distributions (Fig. \ref{fig_RSca}). The inset shows the same functions in a 
semi-logarithm scale, for ages up to $15$. Typical error bars are shown for three 
of the points.}
\label{fig_Mort}
\end{figure}

In Fig. \ref{fig_Mort} these functions are plotted, after having rescaled the age 
distributions. In a linear scale, these functions appear to collapse for young ages, 
and they diverge clearly at the large age end. The inset, on a log-linear scale, 
shows that the mortality functions have the same general behaviour in the small age 
interval shown, but the plot shows an increasing separation between the smaller and 
larger genomes. The collapse is not fully obtained, as can be seen with the help 
of the error bars shown.\\ 

\begin{figure}
\begin{center}
\vspace*{0.8cm}
\resizebox{0.4\textwidth}{!}{\includegraphics{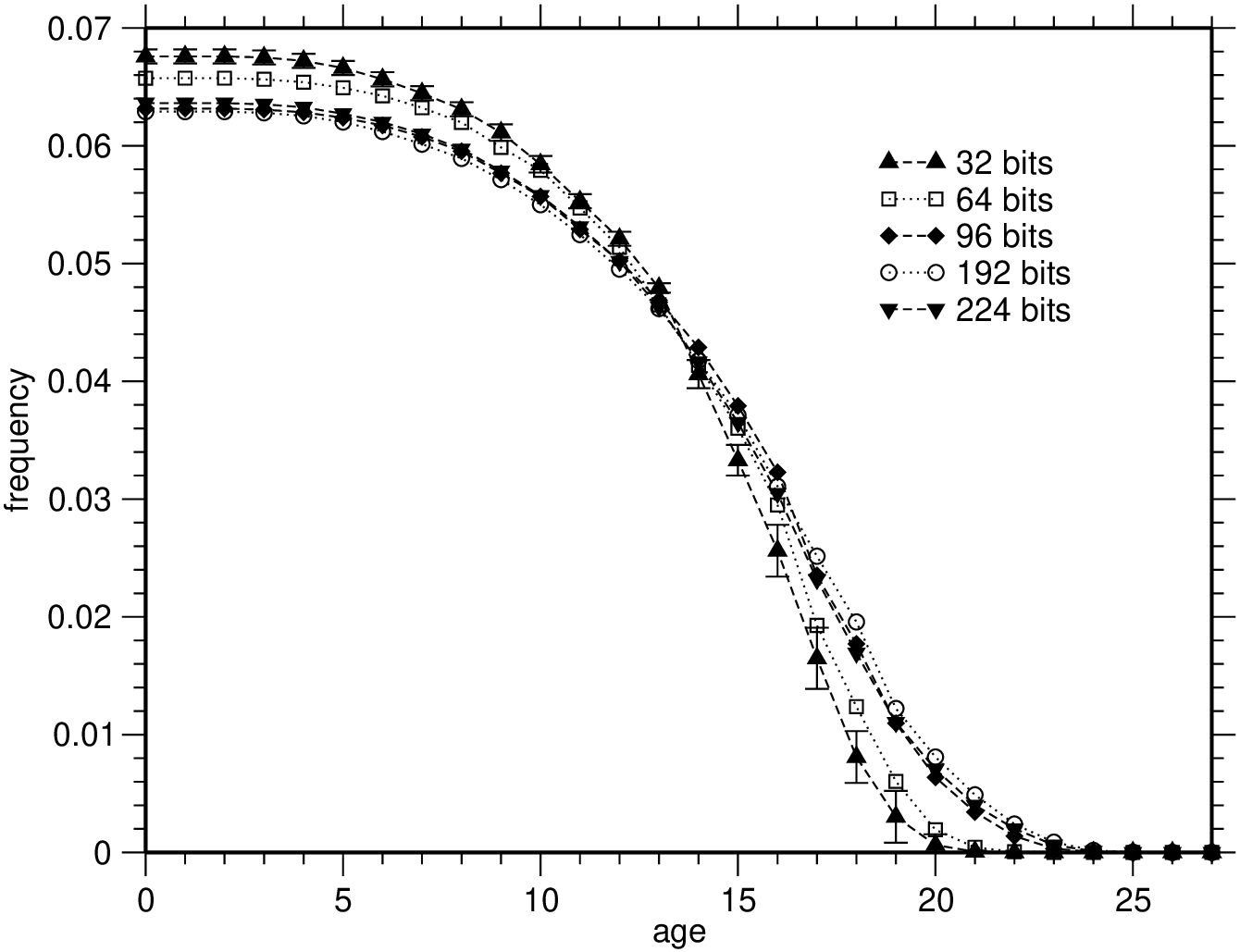} }
\vspace*{0.4cm}
\end{center}
\caption{\small The different simulations when the number of mutations and 
dominant loci are renormalised depending on the string size to keep constant  
their density in the genomes. All the simulations have a duration of 50000 
Monte Carlo steps.}
\label{fig_Norm}
\end{figure}

The slope of the scaling transformation is obviously strongly dependent on the 
values of the simulation parameters. Of particular interest is a choice of these 
parameters that leads to a unit slope. In this case, we may recover the solution of 
the 32-bits model from the others just by rescaling those parameters, which amounts 
to performing a renormalisation. To explore this alternate path, we have focused our 
attention on just two of the parameters, namely the number of dominant loci for the 
harmful allele and the number of mutations added in each generation. The guideline 
here was to keep constant the density of mutations and dominant loci in the genomes, 
independently of their size. We only need to multiply the original values of these 
parameters in the 32-bits model by $i$, the genome size divided by 32. The results 
of this renormalisation procedure are shown in Fig. \ref{fig_Norm}. 

\section{Conclusion}

From the results of our numerical simulations it emerges that, given a Penna model 
with a Verhulst factor acting a single time in each individual's life, the scaling 
laws:

\begin{eqnarray}
32 & \longrightarrow & N = 32 \cdot j
\nonumber\\
t & \longrightarrow & [1 + 0.5 \cdot (j-1)] t+ [0.5 \cdot (j-1)]
\nonumber\\
\rho & \longrightarrow & [1 + 0.5 \cdot (j-1)]^{-1}\rho
\nonumber
\end{eqnarray}

\noindent
where $N$ is the number of bits in the genome and $j$ an integer, lead to age 
distribution functions $(\rho=\rho(t))$ that have similar behaviours, although they 
do not agree quantitatively for all genome sizes. This fact allows one to use any 
genome size in a simulation, if only qualitative features are focused, from which 
the age distribution for all other sizes can be roughly derived. It is also known 
that the situation is no more clearer if the threshold for harmful mutations is 
scaled in proportion to the genome size \cite{private}.

As a final comment, our results seem to indicate that the onset of ageing, usually 
considered as coincident with the minimum reproduction age, is now, for large 
genomes, deferred. The age distributions do show a decreasing trend, starting close 
to the onset of reproduction, but they have very small derivatives - reflected on 
the plateau at small ages for the mortality functions. The lifespan of the 
population increases linearly with genome size, as opposed to being strongly 
dependent only on the minimum reproduction age. The latter prediction is usually 
considered to be a trivial consequence of the mutation accumulation theory on which 
the Penna model is based. These results are somewhat intriguing and deserve further 
investigation.

\section*{Acknowledgements}
The authors wish to thank the anonymous referees, whose comments were instrumental 
for the final form of this paper. We thank the Brazilian agencies CAPES, CNPq, and 
a grant from PRONEX for partial financial support. JSSM acknowledges a special grant 
from FAPERJ.

\begin{table}
\noindent
\caption{Coefficients $c$ and $d$ obtained from regressions of the integral 
function, for $i = 1$ and several values of $j$.}
\begin{tabular}{cccccc}
j & 2 & 3 & 6 & 7 &\\
c & 1.53 & 2.02 & 3.45 & 3.83 &\\
d & 0.51 & 1.06 & 2.58 & 3.12 &\\
\end{tabular}
\label{indices}
\end{table}

\end{multicols}

\end{document}